\documentclass[twocolumn,aps,showpacs,amsmath,amssymb]{revtex4}

\usepackage{graphicx}
\usepackage{dcolumn}
\usepackage{bm}

\begin{document}      

\title{Anomalous high-temperature Hall effect on the triangular lattice in 
Na$_x$CoO$_2$} 
\author{Yayu Wang$^1$, Nyrissa S. Rogado$^2$, R. J. Cava$^{2,3}$, and N. P. Ong$^{1,3}$}      
\affiliation{
$^1$Department of Physics, 
$^2$Department of Chemistry, 
$^3$Princeton Materials Institute,\\
Princeton University, Princeton, New Jersey 08544. 
}

\date{\today}      

\begin{abstract}
The Hall coefficient $R_H$ of $\rm Na_xCoO_2$ ($x = 0.68$) behaves anomalously at high 
temperatures ($T$).  From 200 to 500 K, $R_H$ increases linearly with $T$ to 8 times the 
expected Drude value, with no sign of saturation.  Together with the thermopower $Q$, 
the behavior of $R_H$ provides firm evidence for strong correlation.  We discuss the 
effect of hopping on a triangular lattice and compare $R_H$ with a recent prediction by 
Kumar and Shastry.
\end{abstract}
\maketitle                   

The layered cobaltates are known to exhibit fairly high conductivities for perovskites.  
The discovery by Terasaki, Sasago, and Uchinokura~\cite{Terasaki} of enhanced 
thermopower in the layered Na-doped cobaltate Na$_x$CoO$_2$ provided the first hint that 
these materials may harbor unusual electronic properties stemming from strong Coulomb 
interaction.  Recently,  interest in the cobaltates has escalated with the report of 
Takada \emph{et al.}~\cite{Takada} that superconductivity occurs below $\sim$5 K in the 
oxyhydrate $\rm Na_xCoO_2\cdot yH_2O$.  In addition, Wang \emph{et al.}~\cite{Wang} 
reported that the thermopower $Q$ in $\rm Na_xCoO_2$ ($x = 0.68$) is completely 
suppressed by a 10-Tesla longitudinal magnetic field.  The field suppression of $Q$, 
constituting direct evidence for a large spin-entropy term in the Seebeck coefficient, 
implies that a strong-correlation picture is necessary to describe the electronic 
transport properties.  Superconductivity in $\rm Na_xCoO_2\cdot yH_2O$ has been 
confirmed by several groups~\cite{super}.   

Motivated by the observed field-suppression of $Q$, we have measured in detail the Hall 
effect in $\rm Na_xCoO_2$ ($x\simeq$ 0.68), and found that the dependence of the Hall 
coefficient $R_H$ on temperature $T$ is anomalous.  Above 200 K, the Hall coefficient 
$R_H$ displays a steep, $T$-linear increase with no signs of saturation up to our 
highest $T$ (500 K).  The linear increase continues through a previously unreported weak 
transition at $T_D$ = 430 K.  In addition to the anomalous high-temperature behavior, 
the Hall angle and resistivity are also highly unusual at low-$T$.  

Several groups have proposed phase diagrams for the layered cobaltates based on the 
resonating valence bond (RVB) theory applied to the triangular 
lattice~\cite{Baskaran,Shastry,Lee}.   Independent of our Hall measurements, Kumar and 
Shastry~\cite{Shastry} recently predicted that, on the triangular lattice, the 
high-frequency Hall coefficient $R_H(\omega)$ should be linear in $T$.   We compare our 
experiment with this interesting prediction.

$\rm  Na_xCoO_2$ is comprised of layers of edge-sharing tilted octahedra.  Each 
octahedron is made up of a Co ion surrounded by six O atoms at the vertices.  Within 
each CoO$_2$ layer, the Co ions occupy the sites of a triangular lattice.  Doping is 
accomplished by Na ions which partially occupy the sites of a triangular lattice between 
the CoO$_2$ layers.   With the Na content fixed at $x$, the ratio of Co$^{3+}$ to 
Co$^{4+}$ is $x:(1-x)$.  As common in cobaltates, the splitting between the $e_g$ and 
$t_{2g}$ states in each octahedron is so large that the Co ions are always in their 
low-spin states.  A further trigonal distortion splits the highest lying $t_{2g}$ state 
(called $A_g$) from the remaining 2 $E_g$ states.  In the Co$^{4+}$ ion, the $A_g$ state 
is occupied by one electron, while the $E_g$ states are completely full.  According to 
band-structure calculation~\cite{Singh}, the Fermi level (for $x = 0.5$) falls near the 
top of the band derived from the $t_{2g}$ states.  At $x$ = 0 (all Co ions of formal 
valence Co$^{4+}$), the band formed from the $A_g$ state is half-filled.  Within the 
strong-interaction picture favored by the $Q$-suppression experiment ~\cite{Wang}, this 
half-filled band is a Mott insulator.  Increasing the Na content $x$ is equivalent to 
doping away from the half-filled Mott limit by adding electrons.

\begin{figure}[h]			
\includegraphics[width=8.5cm]{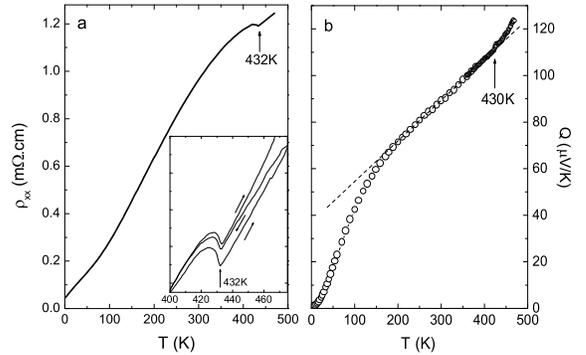}
\caption{\label{rho}  (a) The in-plane resistivity $\rho$ of $\rm Na_xCoO_2$ ($x = 
0.68$).  $\rho$ is linear in $T$ from 2 to 80 K, but has a steeper slope above 100 K.  
The inset shows slight hysteresis in $\rho$ in the vicinity of the transition at $T_D$ = 
430 K.  (b) The in-plane thermopower $Q$ of $\rm Na_xCoO2$.  The anomaly at $T_D$ is 
just resolved in $Q$. 
}
\end{figure}
In our crystals, the Na content is determined~\cite{Wang} by inductive coupled plasma 
analysis to be $x$ = 0.68, which implies that the ratio of $\rm Co^{3+}: Co^{4+}$ is 
$2:1$.  This is confirmed by a plot of the inverse in-plane susceptibility $\chi^{-1}$ 
which displays a Curie-Weiss behavior with a slope consistent with this ratio of 
Co$^{4+}$ ions with $S = \frac12$ (the Weiss temperature is $\theta_W\sim$55 
K)~\cite{Wang}.

As shown in Fig. \ref{rho}, the in-plane resistivity varies linearly from 2 to 100 K 
with a slope $d\rho/dT = 2.45\ \mu\Omega$cm/K about 5 times larger than in 
optimally-doped $\rm YBa_2Cu_3O_7$.  At high $T$, $\rho$ reveals a weakly hysteretic 
transition at $T_D$ = 430 K. Because the transition does not impact qualitatively the 
overall behavior of $\rho$, $Q$ and $R_H$, we associate it with an entropically driven 
order-disorder transition of the Na ions.  The `metallic' profile of $\rho$ suggestive 
of a large Fermi Surface (FS), juxtaposed with a large Curie-Weiss 
susceptibility~\cite{Wang,Ray} indicative of fluctuating local moments with 
antiferromagnetic (AF) coupling, presents an interesting dichotomy if we assume that the 
same electrons are responsible for transport and magnetization.  In fact, the metallic 
profile of $\rho$ masks a set of anomalous transport properties which we describe next.

As noted~\cite{Terasaki}, the thermopower $Q$ is about 10 times larger than predicted by 
the Fermi-liquid expression $Q = (\pi^2/3) (k_B/e)(T/T_F)$ with a Fermi temperature 
$T_F\sim 10^{4}$ K.  A clue to this enhancement derives from the sensitivity of $Q$ to a 
longitudinal magnetic field $\bf H$.  Wang \emph{et al.}~\cite{Wang} found that, below 
20 K, $Q$ is increasingly suppressed by $\bf H$, until at 2 K, it is driven to zero by a 
field of 10 T.  The suppression provides direct evidence that spin entropy contributes a 
dominant fraction of the enhanced $Q$.  The curve of $Q$ vs. $H$ fits closely to a 
simple model of non-interacting spins with $s = \frac12$ and a Lande $g$-factor of 2.2 
$\pm$ 0.1 (Ref. \cite{Wang}).  The plot in Fig. \ref{rho} shows that $Q$ increases with 
nominally constant slope $dQ/dT$ from 200 K to about 450 K, above which the slope 
increases slighly.  The transition at $T_D$ is barely resolved as a weak peak.

\begin{figure}[h]			
\includegraphics[width=6cm]{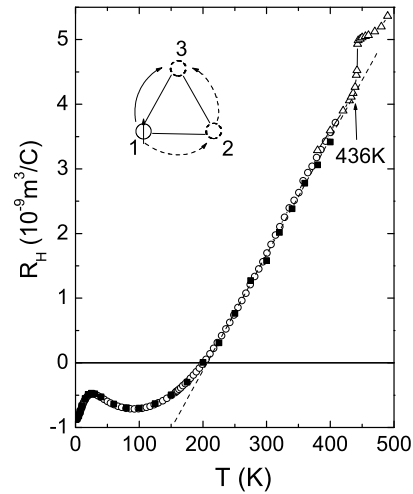}
\caption{\label{Hall}  The $T$ dependence of the Hall coefficient $R_H$ in $\rm 
Na_xCoO_2$ showing anomalous $T$-linear increase between 200 and 500 K.  The open 
circles are measurements using Method 2 on Sample 1.  Solid squares (open triangles) are 
data taken with Method 1 (2) on Sample 2.  The step anomaly at 450 K reflects the effect 
of the disorder transition at $T_D$.  The inset shows the effect of the Peierls phase 
factor on 3 sites.  Interference between the direct process $\hat{t}_{31}$ (solid arrow) 
and the second-order process $\hat{t}_{32}\hat{t}_{21}$ (broken arrows) leads to a 
$\sigma_H\sim  \hat{t}_{13} \hat{t}_{32} \hat{t}_{21}$.
}
\end{figure}
In view of the unusual $Q$, it is natural to investigate other transport properties.  
The Hall coefficient was previously reported~\cite{Terasaki} to be negative and only 
weakly $T$ dependent in the restricted range of $T$ between 4 and 130 K.  When we 
extended the measurements above 200 K, we found the suprisingly strong $T$ dependence 
shown in Fig. \ref{Hall}.  Near 200 K, $R_H$ changes sign and begins a steep increase 
that shows no sign of saturation up to our highest $T\sim$ 500 K, apart from a weak 
anomaly at the transition at $T_D$.  The slope of $R_H$ between 200 and 500 K is nearly 
independent of $T$ away from the anomaly at $T_D$.  The ratio $2:1$ of the formal 
valence charge implies that the carrier density equals 8.65$\times 10^{21}\ {\rm 
cm}^{-3}$, and a Drude value for the Hall coefficient of $|R^0_{H}| = 0.71\ \times 
10^{-9}\ \mathrm{m}^3/C$.  Here, we see that $R_H$ at 500 K is $\sim$8 times larger than 
$R^0_{H}$.

\begin{figure}[h]			
\includegraphics[width=7cm]{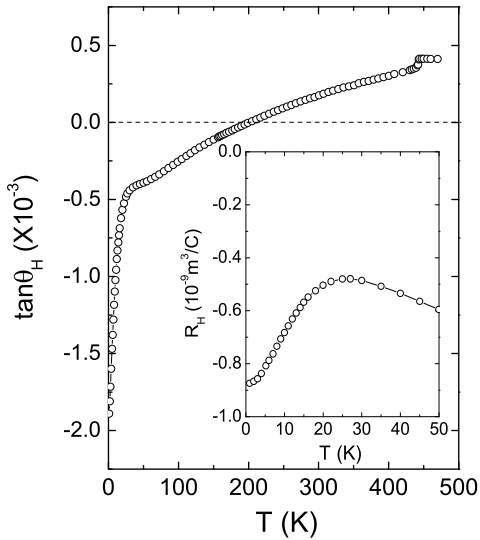}
\caption{\label{tan} The $T$ dependence of the Hall angle $\tan\theta_H$ (measured at 1 
T) in $\rm Na_xCoO_2$.  The profile suggests a crossover to a different scattering 
regime near 25 K.  At low $T$, $|R_H|$ begins to increase rapidly before saturating 
below 3 K (inset). 
}
\end{figure}

Because this Hall profile is so starkly different from the standard $R_H$ profile in 
conventional metals, we have adopted several precautions and checks.  We used two 
methods that complement each others strengths.  Method 1 is the standard practice of 
sweeping $H$ and measuring the Hall voltage $V_H$ at fixed $T$.  At each $T$, $H$ is 
slowly ramped from -14 T to +14 T and then back to -14 T to yield 2 measurements of 
$R_H$.  Method 2 is the technique of Sample \emph{et al.}~\cite{Sample} in which $H$ is 
fixed while $T$ is varied very slowly.  Each measurement of $R_H$ is made by 
electronically toggling between the voltage and current leads to extract the 
antisymmetric component of the resistivity tensor.  Method 1 allows checks for 
non-linearities in $V_H$ vs. $H$, but is error-prone if $T$ drifts significantly during 
the field ramp.  Method 2 provides a high-density trace of $R_H$ vs. $T$ and is robust 
against contact noise, but is unreliable if $R_H$ varies with $H$.  With method 2, $R_H$ 
is seen to have an anomaly at the transition $T_D$, but resumes its $T$-linear trend 
several degrees above $T_D$.  Method 1 reveals that $R_H$ is strictly independent of $H$ 
up to 14 T above 20 K.  Below 20 K, $V_H$ reveals a slight curvature vs. $H$.  As shown 
in Fig. \ref{Hall}, the 2 methods agree well over the whole range of $T$.  

In conventional metals, the empirical rule that $R_H$ must saturate to a constant above 
a `Hall scattering' temperature $\Theta_H$ is nearly universally obeyed in 
non-ferromagnetic metals.  Here $\Theta_H$ is the temperature above which the average 
momentum $\bf q$ of the available scattering excitation exceeds the largest Fermi 
Surface (FS) caliper.  For phonon scattering, $\Theta_H\simeq s\Theta_D$, where $s$ = 
0.2-0.3 and $\Theta_D$ is the Debye temperature.  [In $\rm Na_{0.5}CoO_2$, $\Theta_D$ = 
380 K (Ref. \cite{Ando}), so we would have $\Theta_H\sim$ 100 K if phonon scattering 
were predominant.]  The basis for this rule is clearest in the two-dimensional (2D) FS, 
where the Hall conductivity $\sigma_{H}$ is proportional to the area ${\cal A}_\ell$ 
swept out by the mean-free-path $\vec{\ell}({\bf k})$ as the wavevector $\bf k$ goes 
around the FS~\cite{Ong}.  For $T>\Theta_H$, an electron can always encounter an 
excitation with large enough $|\bf q|$ to scatter it to an arbitrary point on the FS.  
The $\bf k$ dependence of $\ell({\bf k})$ then becomes independent of $T$ apart from an 
overall scale factor.  Hence the \emph{geometric shape} of ${\cal A}_\ell$ is 
$T$-independent, and its magnitude is rigorously proportional to $\ell_{max}^2$, where 
$2\ell_{max}$ is the maximum caliper of the swept area.  As the conductivity $\sigma\sim 
\ell_{max}$, this immediately implies that $R_H$ is independent of $T$.  In the 
published literature of the Hall effect in metals that are non-ferromagnets 
~\cite{Hurd}, this empirical rule is rigorously obeyed (even in systems with both hole 
and electron FS pockets and in $3d$ metals with FS `monsters').  The only known 
exception is the cuprate family.  The strong violation of this empirical rule provided 
an early, conspicuous clue~\cite{Chien} that the charge transport in the cuprates is 
anomalous.  In Fig. \ref{Hall}, the violation is just as striking as in the cuprates, 
but has the opposite trend.  

A different perspective of the anomalous Hall effect is provided by the Hall angle 
$\tan\theta_H = \sigma_H/\sigma$, which is displayed in Fig. \ref{tan}.  In conventional 
metals, $\tan\theta_H$ is proportional to $\ell_{max}$, and must therefore decrease as 
$T$ increases.  Here, we see just the opposite trend.  $\tan\theta_H$ \emph{increases} 
with $T$ right through the zero-crossing at 200 K as well as the disorder transition at 
$T_D$.

\begin{figure}[h]			
\includegraphics[width=8cm]{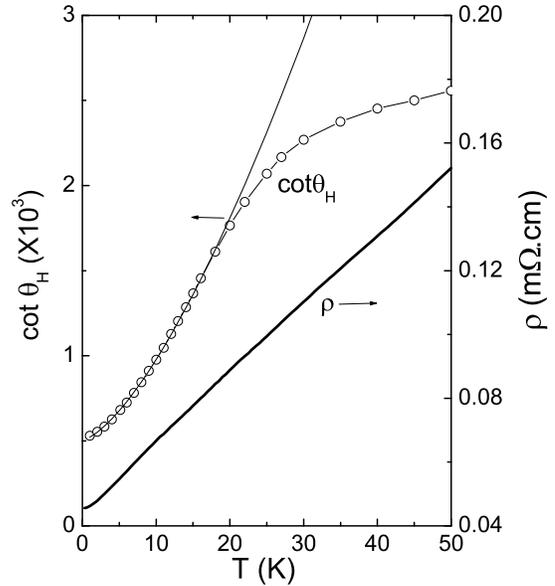}
\caption{\label{lowT}  The $T$ dependence of $\rho$ and $\cot\theta_H$ (measured at 1 T) 
below 50 K.  The fit to $-\cot\theta_H\sim T^n$ (dashed line) shows that $\rho_{xy}$ is 
strongly $T$ dependent because the Hall angle exponent $n = 1.5$ is larger than the 
power-law exponent $n' = 1$ of the transport lifetime.
}
\end{figure}
The sharp change in slope near 25 K in $\tan\theta$ vs. $T$ implies that the Hall 
response crosses over to a distinct regime at low $T$.  As shown in the inset to Fig. 
\ref{tan}, the magnitude of $R_H$ increases steeply below 25 K and saturates to a 
constant only below 3 K.  However, $\rho$ is rigorously $T$-linear from 80 K down to 
$\sim 2$ K as noted above (see Fig. \ref{lowT}).  These profiles of $\rho$ and $R_H$ are 
reminiscent of the corresponding curves in the cuprates (except there the curves extend 
to 500 K or higher and $R_H$ is positive).  Following the analysis of Chien \emph{et 
al.}~\cite{Chien} in the cuprates, we find that $\cot\theta_H$ is well-fitted by the 
expression $a + bT^n$ with $n= 1.5$ (dashed line in Fig. \ref{lowT}).  This implies that 
the strong $T$ dependence of $|R_H|$ below 25 K results from distinct power-law 
exponents for the Hall-angle lifetime ($n = 1.5$) and the transport lifetime obtained 
from $\rho$ $(n' = 1)$.  Interestingly, the temperature scale over which this holds is 
about 10-20 times smaller than in the cuprates.  This analysis suggests that, below 25 
K, the carriers in $\rm Na_xCoO_2$ enter into a strong-correlation regime that involves 
the same `strange-metal' physics as observed in the cuprates.   The reduction of the 
Hall-angle exponent from 2 to 1.5 also matches well the observed reduction from 
optimally doped to overdoped cuprates.  

We return to the high-temperature behavior of $R_H$.  The Hall effect problem in 
cuprates has motivated many theoretical investigations of $R_H$ in the $t$-$J$ model.  
Recently, Kumar and Shastry (KS) applied to $\rm Na_xCoO_2$ a series expansion 
technique~\cite{SSS} to calculate the high-frequency limit of the Hall coefficient 
$R^*_H$ (the small parameter is $\beta t$ where $t$ is the hopping amplitude and $\beta 
= 1/k_BT$).  For a triangular lattice, KS obtain~\cite{Shastry,SSS}
\begin{equation}
R^*_H  = -\frac{a^2d}{8|e|} \frac{k_BT}{t}\frac{(1+\delta)}{\delta(1-\delta)},
\label{RH}
\end{equation}
where $a$ is the Co-Co bond length, $d$ the interlayer spacing, and $\delta$ the doping 
away from half-filling.  

In Fig. \ref{Hall}, $R_H$ is $T$-linear between 200 and 500 K.  If we fit this linear 
portion to Eq. \ref{RH}, we find that $t$ = 25 K, consistent with $\beta t < 1$.  
However, it is not obvious \emph{a priori} that the prediction for $R^*_H$ can be 
compared meaningfully with the dc data in Fig. \ref{Hall} (moreover, the theory does not 
predict the weaker dependence on $T$ below 100 K).  It is clearly important to verify 
theoretically whether the domain of validity of Eq. \ref{RH} extends to the dc regime in 
the $t$-$J$ model.

The singular behavior of $R_H$ is intimately related to charge transport on the 
triangular lattice.   It is helpful to recall electrons hopping on a lattice (Fig. 
\ref{Hall}, inset).  The Hall current arises from the Peierls phase factor ${\rm 
e}^{i\alpha}$ acquired around the minimal closed loop, where $\alpha = e\phi/\hbar$ is 
proportional to the flux $\phi$ piercing the loop.  For the 3-site problem (site indices 
1,2,3), Holstein~\cite{Holstein,Friedman} computed the transition rate for hopping from 
site 1 to 3 in a field, and found that the direct, first-order hopping amplitude $ 
\hat{t}_{31}$ interferes with the second-order amplitude $ \hat{t}_{32} \hat{t}_{21}$ 
(with $ \hat{t}_{ij}$ the complex hopping amplitude between sites $i$, $j$ in a field), 
to yield a giant enhancement of the Hall mobility (Fig. \ref{Hall}, 
inset)~\cite{Holstein}, so that $\sigma_H\sim \hat{t}_{13} \hat{t}_{32} \hat{t}_{21}\sim 
t^3{\rm e^{i\alpha}}$.  However, this non-interacting, hopping model obviously does not 
describe the physics here.

To get a $T$-linear $R_H$, we need to consider strong correlation.  In the $t$-$J$ model 
calculation of KS, transport around a loop on the triangular lattice gives likewise 
$\sigma_H\sim (\beta t)^3\sin\alpha$~\cite{Shastry}.  Since $\sigma\sim (\beta t)^2$, we 
have $R_H\sim (\beta t)^{-1}$, which implies a $T$-linear increase.  In the 
strong-correlation problem, the singular behavior in Eq. \ref{RH} is a consequence of 
transport on the triangular lattice.  

The accumulated evidence increasingly support the strong-interaction picture in the 
layered cobaltate.  Initially, observations of an enhanced $Q$~\cite{Terasaki} and a 
Curie-Weiss susceptibility~\cite{Ray} in the presence of a `metallic' $\rho$ provided 
early hints of strong correlation.  Later, direct evidence for dominant spin entropy 
contribution to the thermopower came from the effect of $H$ on $Q$ ~\cite{Wang}.  The 
steep increase in $R_H$ at high temperatures, as well as the cuprate-like behavior of 
$R_H$ and $\rho$ below 25 K, reported here provide further evidence for strange-metal 
behavior.  Both the field suppression of $Q$ and the Hall profile in Fig. \ref{Hall} 
provide striking examples of how strong interaction can change \emph{qualitatively} 
electronic transport behavior from conventional behavior even at temperatures up to 500 
K.  Hopefully, a systematic search may uncover more examples of $T$-linear $R_H$ in 
layered perovskites with triangular lattices.

We thank B. S. Shastry for enlightening discussions, and acknowledge support from a 
MRSEC grant (DMR 0213706) from the National Science Foundation.


\begin{thebibliography}{99}
%
\bibitem{Terasaki} I. Terasaki, Y. Sasago, and K. Uchinokura, Phys. Rev. B {\bf 56}, 
R12685 (1997).
\bibitem{Takada} K. Takada \emph{et al.}, Nature {\bf 422}, 53 (2003).
\bibitem{Wang} Yayu Wang, Nyrissa S. Rogado, R. J. Cava, and N. P. Ong, Nature, \emph{in 
press}.
\bibitem{Singh} D. J. Singh,  Phys. Rev. B {\bf 61}, 13397 (2000).
\bibitem{super} M.L. Foo \emph{et al.}, cond-mat/0304464;  B. Lorenz \emph{et al.}, 
cond-mat/0304537;  F. Rivadulla \emph{et al.}, cond-mat/0304455.
\bibitem{Ray} R. Ray, A. Ghoshray, K. Ghoshray and S. Nakamura, Phys. Rev. B {\bf 59}, 
9454 (1999).
\bibitem{Ando} Y. Ando \emph{et al.}, Phys. Rev. B {\bf 60}, 10580 (1999).
\bibitem{Baskaran} G. Baskaran, cond-mat/0303649.
\bibitem{Shastry} Brijesh Kumar and B. Sriram Shastry, cond-mat/0304210.
\bibitem{Lee} Qiang-Hua Wang, Dung-Hai Lee and Patrick A. Lee, cond-mat/0304377.
\bibitem{SSS} B. S. Shastry, B. I. Shraiman and R. R. P. Singh, Phys. Rev. Lett. {\bf 
70}, 2004 (1993).
\bibitem{Sample} H.H. Sample \emph{et al.}, Jnl. Appl. Phys. {\bf 61}, 1079 (1987).
\bibitem{Hurd} \emph{See for e.g.}, C. M. Hurd, \emph{The Hall Effect in Metals and 
Alloys} (Plenum, New York, 1972). 
\bibitem{Ong} N. P. Ong, Phys. Rev. B {\bf 43}, 193 (1991).
\bibitem{Chien} T. R. Chien, Z. Z. Wang and N. P. Ong, Phys. Rev. Lett. {\bf 67}, 2088 
(1991).
\bibitem{Holstein} T. Holstein, Phys. Rev. {\bf 124}, 1329 (1961).
\bibitem{Friedman} L. Friedman and T. Holstein, Ann. Phys. (N.Y.) {\bf 21}, 494 (1963).





%

\end{thebibliography}
\end{document}